\documentclass[a4paper, amsfonts, amssymb, amsmath, reprint, showkeys, nofootinbib, twoside]{revtex4-1}
\usepackage[english]{babel}
\usepackage[utf8]{inputenc}
\usepackage[colorinlistoftodos, color=green!40, prependcaption]{todonotes}
\usepackage{amsthm}
\usepackage{mathtools}
\usepackage{physics}
\usepackage{xcolor}
\usepackage{graphicx}
\usepackage[left=23mm,right=13mm,top=35mm,columnsep=15pt]{geometry} 
\usepackage{adjustbox}
\usepackage{placeins}
\usepackage[T1]{fontenc}
\usepackage{lipsum}
\usepackage{csquotes}
\usepackage[pdftex, pdftitle={Article}, pdfauthor={Author}]{hyperref} 
\begin{document}
\title{Computational study of Adhesion and Friction Behavior of Crosslinked Polymer Network}

\author{Ajay Kumar$^{}$}

\author{Manoj Kumar Maurya$^{}$}
\email[Correspondence email address: ]{manojmaurya647@gmail.com}

\author{Manjesh K. Singh$^{}$}
\email[Correspondence email address: ]{manjesh@iitk.ac.in}

\affiliation{$^{}$~Department of Mechanical Engineering, Indian Institute of Technology Kanpur, Kanpur UP 208016, India.}


\begin{abstract}
In this study, we have utilized a molecular dynamics simulation approach to understand adhesion and friction behaviour of crosslinked polymer networks. We have used breakable quartic bond to model crosslinked polymers. We explored the structural characteristics and evaluated the coefficient of friction (CoF) as a function of crosslinked monomer fraction (cross-linking bond density) in four-fold cross-linked polymer networks. To estimate CoF, a rigid indenter was inserted to different depths of indentation. Subsequently a constant sliding speed was applied while keeping the depths of indentation fixed. Normal and friction forces were calculated at each depth to estimate CoF through a linear curve-fitting. For adhesion studies, using the force vs. displacement curve we quantified adhesion through the forces during the separation of rigid indenter from surface of crosslinked polymeric materials while unloading after indentation into the sample. The results indicate that as the fraction of crosslinked monomers increases, the stiffness of the crosslinked network increases, while the force of adhesion and CoF decrease. Additionally, increasing the depth of indentation during friction leads to higher frictional forces.
\end{abstract}


\maketitle
\section{Introduction}
\label{sec: Introduction}

In recent years, researchers have shown growing interest in a new class of polymers called crosslinked polymers\cite{mauryaHCP, mauryaWCP, chen2020intramolecularly, nielsen1969cross, shen2018effects, song2020high}. These are different from regular or linear polymers\cite{nielsen1969cross}, in crosslinked polymers each monomer in the structure can form multiple bonds with neighboring monomers\cite{mukherji2009mechanical,mauryaHCP, mauryaWCP}. This creates a three-dimensional network. Crosslinking changes the physical and chemical properties of the polymer\cite{mane2015effect, kolhe2022review}. Compared to normal polymers, crosslinked polymers usually have better thermal stability\cite{li2001thermal, xiao2019situ, levchik1999correlation, uhl2001thermal}, higher resistance to chemicals, good elasticity and mechanical properties\cite{mauryaHCP}. One important feature of crosslinked polymers is their crosslinking density (degree of crosslinking), which significantly affects the material performance\cite{song2020high, nielsen1969cross}. And the degree of cross-linking can vary, from elastomers, which are weakly cross-linked and flexible\cite{zhang2023hybrid, zentel1986liquid}, to epoxies, which are highly cross-linked and rigid\cite{tsige2004effect, aramoon2016coarse, seo2019development, duvsek1984cross}. Highly cross-linked polymers are widely utilized in various applications, including foamed structures, composite matrices, adhesives, and electronic packaging~\cite{thermoset, quratTI, chen2003new,ain2021overview}. However, weakly cross-linked polymer networks (WCPs) provide greater flexibility and conformational freedom for polymer chains due to their less rigid interconnections~\cite{barbero2022cross,perrot2021use}.
As a result, WCPs are used as artificial antibodies and enzymes.~\cite{weil20jacs}, drug delivery Systems, tissue engineering scaffolds and bioadhesive.
Synthetic polymers were first recognized by Hermann Staudinger in the 1920s. Polymers are essentially materials composed of long chains of repeating monomer units that are connected by covalent bonds.~\cite {stuart2010emerging}. These polymers have unique characteristics at different lengths, times, and energy scales, and each contributes uniquely to the overall behavior and characteristics of the polymer chain~\cite{stuart2010emerging, DMCMMKK2020}. These multiscale phenomena are essential in determining polymeric materials structural integrity, flexibility, and functional performance, from molecular-level interactions to bulk mechanical responses. The flexible nature of polymers plays a very important role in enabling the design of advanced functional materials with tunable properties\cite{zhao2020elastin, mukherji2019soft}. Polymers belong to the class of soft matter\cite{doi2013soft, yakhmi2011soft, morozova2019surface, genzer2008surface} and their properties are governed primarily by weak van der Waals interactions, which are in order of $k_{\rm B}T$ at ambient temperature. Here, $k_{\rm  B}$ is the Boltzmann constant and $T$ is temperature~\cite{stuart2010emerging, DMCMMKK2020,dai2019entropic}. Due to the weak interactions, polymers show large conformational and compositional fluctuations, making them highly suitable for molding into various shapes\cite{linse2010polymer}.

Indentation experiments are particularly useful for measuring the hardness and modulus of materials\cite{oliver1992improved}.
This technique is particularly useful for cross-linked polymers\cite{mauryaHCP, mauryaWCP, kumar2025adhesion, kumar2025controlling}, as their unique network structure significantly affects their mechanical behavior. By applying indentation techniques, we can gain insights into the hardness and deformation characteristics of cross-linked polymers, providing important information on their overall material behavior and durability. In the past, indentation-based work was mainly focused on polymeric systems like hydrogels, microgels, and elastomers. Still, in recent years, researchers have focused on other polymeric materials~\cite{singh2018combined, muser2019modeling, mathesan2016molecular}.
In recent research, the mechanical properties of highly cross-linked polymers (HCPs) and weakly cross-linked polymers (WCPs) were examined using computational indentation with implicit indenters using a molecular dynamics simulation approach\cite{mauryaHCP, mauryaWCP}. The findings revealed that cross-linked polymer networks exhibit distinct behavior compared to linear polymers, demonstrated stiffening upon indentation and increased stiffness with higher cross-linking density~\cite{mauryaHCP, mauryaWCP}. However, to the best of our knowledge, there are no existing studies on crosslinked polymers that investigate the mechanisms of adhesion using an explicit rigid indenter.
In this study, we used an explicit rigid spherical indenter, which allows monomer-level resolution due to localized deformation and provides detailed insights into the microscopic structure at small scales. Tribology involves the study of surfaces in relative motion, focusing on phenomena such as friction, material wear, scratching, and rubbing~\cite{ain2024tribological, kumar2024tribological, kumar2025nano, dubey2024comparative}. More precisely, it is the science and technology concerned with interacting surfaces in motion, including efforts to minimize the costs associated with friction and wear~\cite{bhushan2003introduction}. The tribological behavior of cross-linked polymer networks remains largely unexplored, with limited studies employing molecular dynamics simulations. This research gap constrains the broader application of cross-linked polymeric materials. Müser et al. conducted two-dimensional simulations using a sinusoidal indenter, demonstrating a wave vector-dependent effective modulus influenced by mechanical pressure, cross-linking density, and solvent quality~\cite{muser2019modeling}.


Polymer adhesives are also replacing traditional joining methods such as soldering, bolting, and screwing, particularly in aerospace and automotive engineering, because they are lighter, generate low-stress concentrations, and are more resistant to fatigue~\cite{baldan2004adhesively}. The adhesion characteristics, such as the pull-off force or separation energy, can be determined by analyzing the force-displacement analysis, obtained from the indentation simulations~\cite{greenwood1981mechanics, hui1998contact}. 
Another more robust method would be to analyze the results and quantify the adhesion energy of the encountering surfaces with particular adhesion theories. The  Johnson–Kendall–Roberts (JKR) model \cite{JKR} and Derjaguin–Muller–Toporov (DMT) model \cite{DMT} are extensively used models. Tabor \cite{tabor1977surface} and Maugis \cite{maugis1992adhesion} parameters are used for finding the behavior of materials between the JKR and DMT extremes. A reduction in indenter size corresponds to decreased indentation force and increased adhesion forces during the indentation experiment~\cite{sirghi2006adhesion}, which significantly affects the adhesion performance of the polymeric system~\cite{lee2017effect}.

This study explores the mechanics of cross-linked polymer networks with a broad range of capabilities using large-scale molecular dynamics simulations of a generic model and computational indentation using an explicit rigid spherical indenter. Our model is helpful for a qualitative description, not quantitative, for any chemical-specific systems. We aim to fill existing research gaps by consistently determining the adhesion and coefficient of friction of cross-linked polymer networks with varying fractions of crosslinked monomers.\\ 
The remainder of the paper is organized as follows: In Sec.~\ref{sec: method}, we outline our model and methodology. Sec.~\ref{sec: Result and discussions} present the results and discussions. Finally, the conclusions are drawn in Sec.~\ref{sec: Conclusion}.

\section{Model and Approach}
\label{sec: method}

In this study, we opted for a tetrafunctional crosslinked network\cite{mauryaHCP},
where tetra refers to the maximum number of $4$ bonds that a monomer can form with its adjacent monomers. Here, the systems consist of $N_{\rm c} = 2520$ polymer chains, each chain having a degree of polymerization $N_{\ell} = 100$. The polymer chains are distributed within the cubical box, maintaining the number density $\rho_{\rm m} = 0.85~\sigma^{-3}$. Moreover, we also developed the crosslinked systems that depend on the fraction of crosslinked monomers ($x_{\rm f}$), which represents the functionalization of monomers in the chain to enable the crosslink formation with adjacent monomers. 

In this study, we developed three crosslinked polymer network systems with varying fractions of crosslinked monomers ($x_{\rm f}$), where monomers are functionalized at different intervals corresponding to $x_{\rm f} = 1.0$ (every monomer), $x_{\rm f} = 0.50$ (every second monomer), and $x_{\rm f} = 0.20$ (every fifth monomer) along each polymer chain. For this study, the LAMMPS (Large-scale Atomic/Molecular Massively Parallel Simulator) molecular dynamics package is used to perform the simulations~\cite{plimpton1995fast, singh2015polymer, singh2016effect, singh2018combined, mauryaWCP}.


\subsection{Interatomic force fields}
For this work, we employed a generic bead-spring model, where each bead represents a certain number of atomistic monomers (one or more monomers) and the size roughly corresponding to the Kuhn length. The degree of coarse-grain is controlled by the bond length and bending stiffness. Interactions between the non-bonded monomers are modeled using the Lennard-Jones (LJ) 12/6 potential $u_{\rm nb}(r)$\cite{mauryaWCP} as described in Eq.~\ref{eq: LJ potential}.

\begin{equation}
\label{eq: LJ potential}
u_{\rm nb}(r) =
\begin{cases}
 4\epsilon \left[\left(\frac {\sigma} {r}\right)^{12} - \left(\frac {\sigma}{r}\right)^6 \right]-u_{\rm cut}, & \hspace{1.5cm} r \leq r_{c} \\
 0, & \hspace{1.5cm} r > r_{c}
\end{cases}
\end{equation}

Here, $r$ represents the distance between two monomers, $r_{c}$ represents the cut-off distance ($r_{c}=2.5\sigma$), $\sigma$ and $\epsilon$ are the LJ units of distance and energy, respectively, and mass of the monomers represented by $m$, a unit of time represented by
$\tau = \sigma \sqrt{m/\epsilon}$ and unit of force $F$ is represented by $k_{\rm B}T/\sigma$. Table~\ref{tab: LJ to real unit} shows the conversion of the LJ reduced units into real units for typical hydrocarbons.

\begin{table}[]
\centering
\label{tab: LJ to real unit}
\begin{center}
 \begin{tabular}{|c|c|c|}
\hline
{\color{white}}&&\\

Physical Quantity & L-J Reduced unit &  Real Units\\
\hline
\hline
Length ($r$) & $1~\sigma$ &  0.5~nm \\
\hline
Energy ($E$) & 1~$\epsilon$ & 30~meV  \\
\hline
Time ($t$) & 1~$\tau = \sigma \sqrt{m/\epsilon} $&3.0~ps  \\
\hline
Temperature ($T$) & 1~$\epsilon /k_{\mathrm{B}}$ & 300 K \\
\hline
Pressure ($P$) & $ 1~k_{\mathrm{B}}T/\sigma^{3}$ &  40~MPa \\
\hline
\end{tabular}
\end{center}
\caption{This table shows the typical values for converting LJ units to real-world quantities in hydrocarbon systems~\cite{kremer1990dynamics}.}
\label{tab: LJ to real unit}
\end{table}

In this methodology we used the velocity varlet algorithm to integrate the equation of motion with a time step of $0.005\tau$ and used Langevin thermostat with a damping coefficient of $\gamma = 1.0 \tau^{-1}$ to maintain the system temperature $T = 1.0 \epsilon/k_{\rm B}$ which is generally higher than the glass transition temperature of the system $T_{\rm g}  \approx 0.4 \epsilon/k_{\rm B}$~\cite{stevens2001interfacial,MukherjiPRE2008}.\\

To model the crosslinked polymers network, we used bonded interaction potentials $u_{\rm b}(r)$, Finite extensible nonlinear elastic (FENE)\cite{kremer1990dynamics, singh2015polymer} and breakable quartic potential~\cite{DMPRM21, MukherjiPRE2009}.


\subsection{Cross-linking/bond formation}

During the formation of crosslinked polymer networks with different fractions of crosslinking monomers ($x_{\rm f}$), the FENE potential is used, as described in Eq.~\ref{eq: fene potential}, where the bond between two monomers is defined by a combination of the repulsive (${r < 2^{1/6}\sigma}$) Lennard-Jones (LJ/12-6) potential and nonlinear logarithmic term:

\begin{multline}
	\label{eq: fene potential}
u_{\rm FENE}(r)=4\epsilon\!\left[\left(\frac{\sigma}{r}\right)^{12}
-\left(\frac{\sigma}{r}\right)^6+\tfrac14\right] \\
-\tfrac12 k R_{0}^{2}\ln\!\left(1-\left(\tfrac{r}{R_{0}}\right)^2\right)
\end{multline}

Where, $k= 30 k_{\rm B}T/\sigma^2$ and cutoff distance $R_{\circ} = 1.5\sigma$.\\

Before curing the cross-linked network, the system consists of randomly distributed LJ particles within the cubical box bounded by two repulsive walls along the $z$ axis, and periodic boundary conditions were applied in both the $x$ and $y$ directions.
Using the protocol proposed earlier \cite{MukherjiPRE2009, DMPRM21}, we find that a bond is formed between two monomers if:

\begin{enumerate}

    \item The distance between the two adjacent monomers is less than $1.1 \sigma$.
    
    \item  Monomers are allowed to form a maximum of four bonds with neighboring monomers in the case of a tetrafunctional network.
      
    \item The setting for crosslinked monomers fraction must have a value between $0$ and $1.0$. A randomly generated number between $0$ and $1.0$ should be less than $0.05$ (bond formation probability). 
\end{enumerate}

With the decided conditions, the network was cured with microcanonical ensembles for the time of $\tau = 5\times 10^{3}\tau$, within this curing time, curing of the cross-linked network achieved more than $90\%$ \cite{mauryaHCP, mauryaWCP}. System equilibrated after curing with isothermal-isobaric $(NPT)$ ensemble for the time of $t = 2.55\times 10^{3}\tau$ at zero pressure using Nos\'e{-}Hoover barostat with damping parameter $\gamma_{\rm p} = 0.5\tau$.\\

\subsection{Computational indentation and friction}

In this study, for computational indentation and friction, we used quartic potential in combination with purely repulsive Lennard-Jones (LJ/12-6) potential, as mentioned in Eq.~\ref{eq: quartic}, which allows modeling of bond breaking when the distance between two bonded monomers exceeds $R_{\rm c}$.

\begin{equation}
\label{eq: quartic}
\begin{split}
u_{\rm q}(r) = & \; 4\epsilon \left[\left(\frac{\sigma}{r}\right)^{12} 
- \left(\frac{\sigma}{r}\right)^6 + \frac{1}{4}\right] \\
& + k(r-R_{\rm c})^{2}(r-R_{\rm c}-B_{1})(r-R_{\rm c}-B_{2}) 
+ u_{\rm o}
\end{split}
\end{equation}

\begin{figure*}[ptb]
\hfill
\includegraphics[width=1.0\textwidth,angle=0]{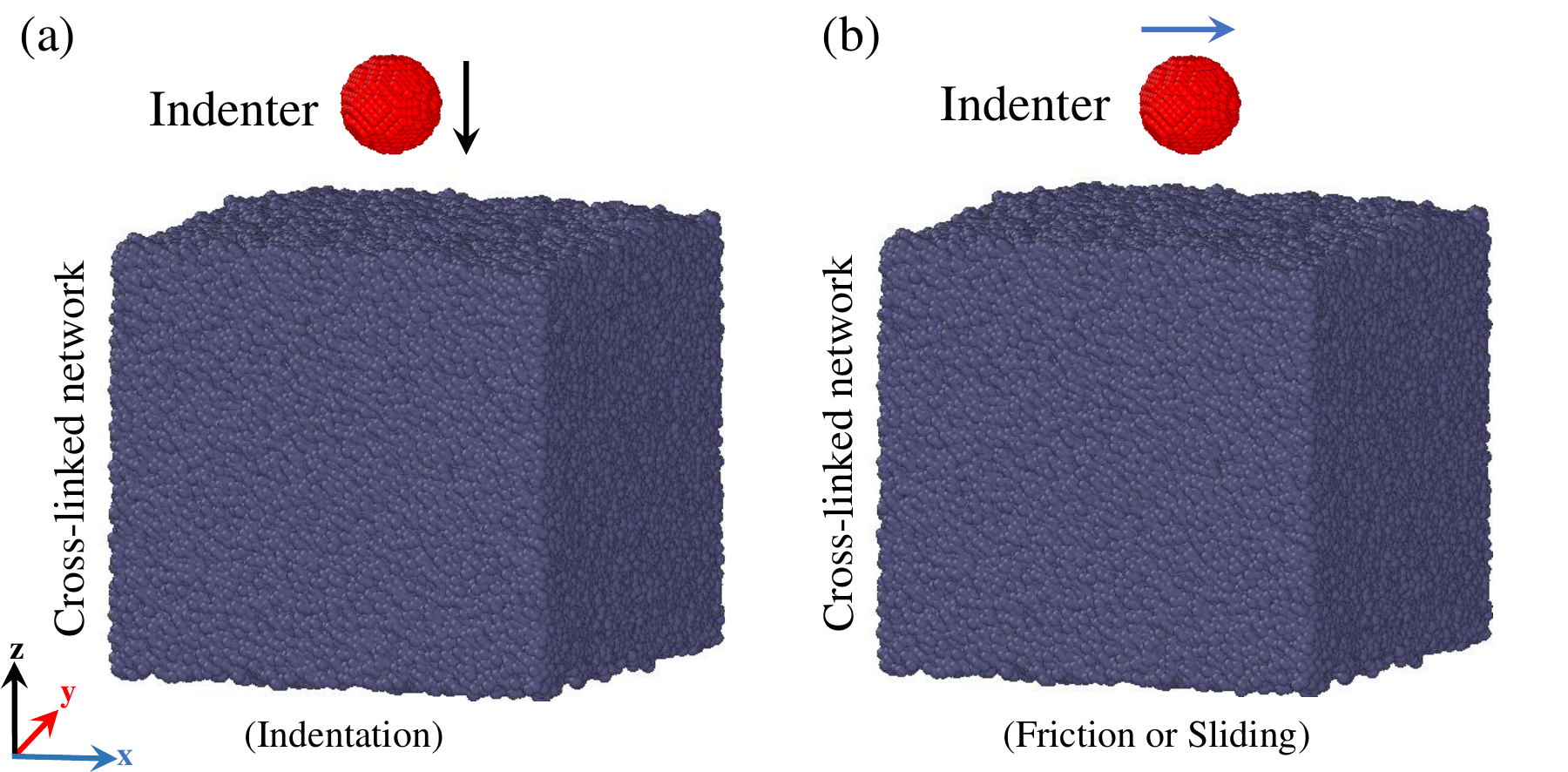}
\caption{Schematic illustration of crosslinked network system with monomers fraction $x_{\rm f} = 1.0$, explicit spherical rigid indenter (red) with radius $r_{\rm ind} = 8.0\sigma$ and substrate (dark blue). (a) The indentation was carried out in the $-z$ direction. (b) indenter slides in $+x$ direction. The simulation snapshot is rendered using OVITO~\cite{ovito}}
\label{fig: Ind_fric_snapshot}
\end{figure*}

Here, $k = 2351 {k_{\rm B}T}/{\sigma^4}$, $R_{\rm c} = 1.5 \sigma$, $B_{\rm 1} = 0 \sigma$, $B_{\rm 2} = -0.7425 \sigma$,
and $u_{\rm o} = 92.744 k_{\rm B}T$.
For the force response,  we used a spherical rigid explicit indenter with a radius ($r_{\rm ind}$) of $8.0~\sigma$. Indentation was performed along the $-z$ direction, as shown in Fig.~\ref{fig: Ind_fric_snapshot}(a), while friction was studied as the indenter slide over the substrate in the $+x$ direction, as represented in Fig.~\ref{fig: Ind_fric_snapshot}(b).


\subsection{Force on indenter}

The total force exerted by the indenter on the substrate originates from the sum of all interatomic interactions between the constituent atoms of the two bodies. If $N_{\rm i}$ represents the number of atoms in the indenter and $N_{\rm s}$ denotes the number of atoms in the substrate, the net force applied by the indenter can be expressed as the sum of the pairwise interaction forces as described in Eq.~\ref{eq: Foece_on_indenter}

\begin{equation}
\label{eq: Foece_on_indenter}
	F_{\rm indenter}=\sum_{i=1}^{N_i}\sum_{j=1}^{N_s} F_{ij}	
\end{equation}

During indentation and friction simulations, the resulting total force is typically resolved in the direction normal to the substrate surface (e.g., z-direction), while the friction force resolutions are carried out in the x-direction. The force exerted on the substrate is equal in magnitude and opposite in direction to that on the indenter:

\begin{equation}
\label{eq: substrate and indenter reactions force}
	F_{\rm substarte}=-	F_{\rm indenter}
\end{equation}
\section{Results and Discussions}
\label{sec: Result and discussions}


\subsection{Indentation response}
\label{sec: indentation response}

\begin{figure*}[ptb]  
\hfill
\centering
\includegraphics[width=1.0\textwidth,angle=0]{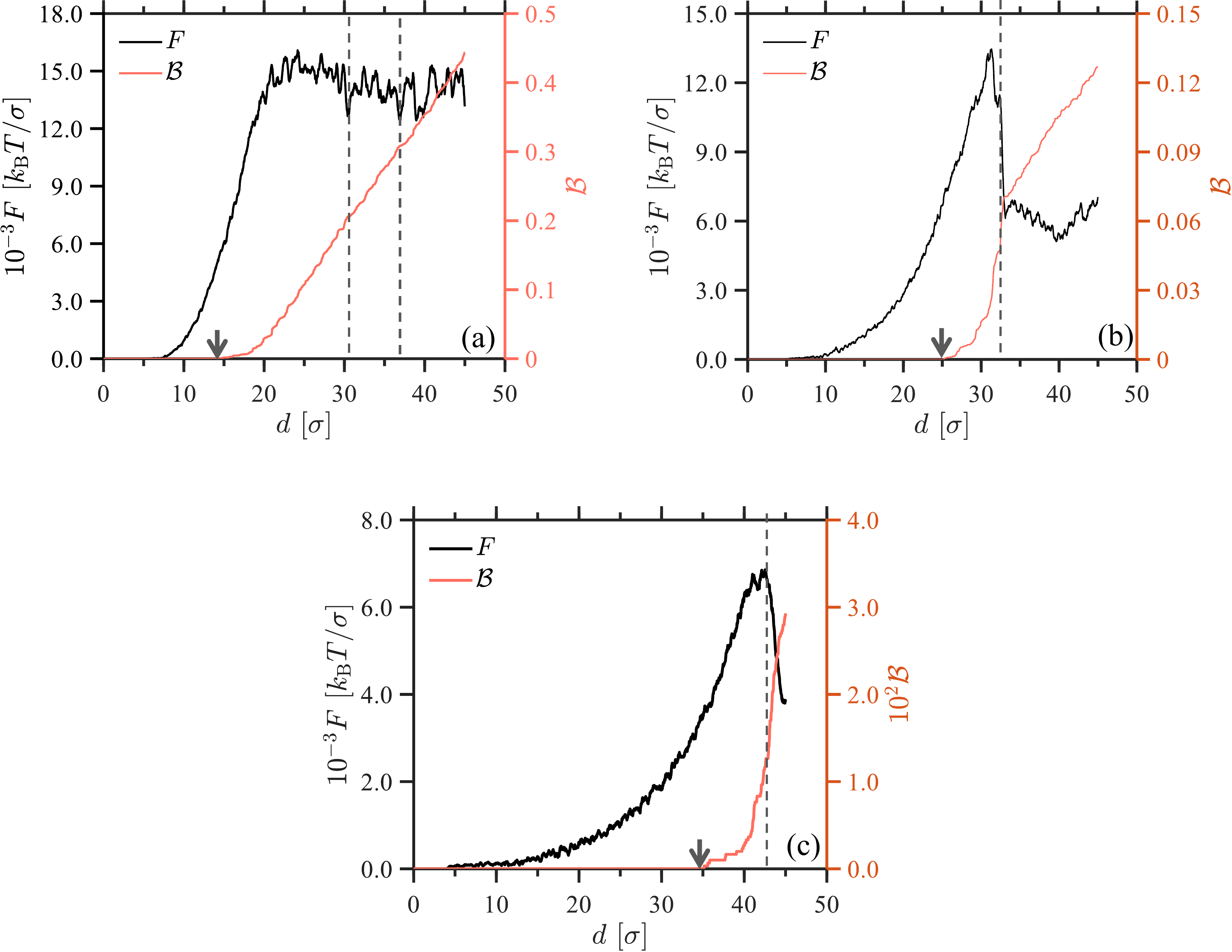}
\caption{Loading Force ($F$) and corresponding bond-breaking percentage ($\mathcal{B}$) 
as a function of indentation depth ($d$). Data are shown for tetrafunctional networks 
with different fractions of crosslinked monomers (a) $x_{\rm f} = 1.0$, 
(b) $x_{\rm f} = 0.5$, and (c) $x_{\rm f} = 0.2$. The color codes of the data sets 
correspond to their respective y-axes. The indenter radius was $r_{\rm ind} = 8.0\sigma$ 
and the approach velocity of the indenter was $v=0.005\sigma/\tau$. 
Vertical dashed lines highlight major force drops ($\Delta F$). 
The indenter tip starts at a certain distance from the upper surface of the substrate, 
and arrows indicate the locations where bond breaking begins.}
\label{fig: Tetra_tri_loading_bond_breaking}
\end{figure*}

An indentation test was conducted to evaluate the mechanical response of crosslinked polymer networks with different fractions of crosslinked monomers ($x_{\rm f}$). The results are presented in Fig.~\ref{fig: Tetra_tri_loading_bond_breaking}(a–c), where the indentation force ($F$) is plotted against the indentation depth ($d$). The graphs show that the loading force increases while the depth of indentation increases, which indicates a stiffer response with deeper penetration. However, at certain points during indentation, there is a sudden drop in the loading force ($\delta F$) observed. This force drop corresponds to bond-breaking that happens within the crosslinked network structures, which are highlighted with dashed lines (gray color) in Fig.~\ref{fig: Tetra_tri_loading_bond_breaking}(a–c). During indentation, the applied stress exceeds the bond strength, causing microscopic failure within the material.

To better understand these force fluctuations $\delta F$, these force drops are considered significant when their magnitudes exceed the fluctuation errors observed in the elastic regime prior to the onset of bond breaking (indicated by the downward arrow in the figures (a-c)). We quantified the percentage of broken bonds ($\mathcal{B}$) as a function of indentation depth (see Fig.~\ref{fig: Tetra_tri_loading_bond_breaking}(a–c) with red color on the right y-axis). These force drops are attributed to the redistribution of energy and atomic rearrangements, which are characteristic of the mechanical behavior seen in glassy materials; however, it is important to highlight here that our system is not glassy. This atomic rearrangement during deformation alters the local structure of the material, eventually leading to a sudden drop in force \cite{falk1998dynamics}. For the same indentation depth, bond breaking began earlier in the crosslinked network with $x_{\rm f} = 1.0$ (highly crosslinked network), as indicated by the downward arrows in the corresponding Fig.~\ref{fig: Tetra_tri_loading_bond_breaking}(a–c). Network with higher $x_{\rm f}$ (crosslinking bond density), resulting in a more rigid structure. The increased rigidity reduces the network flexibility, limiting its ability to distribute the applied stress evenly while indentation. As a result, the stress becomes concentrated in certain areas, causing bonds to break earlier. The load-bearing capacity of the crosslinked network decreases with a reduction in $x_{\rm f}$, which is governed by the fraction of crosslinked monomers.For the same depth of indentation, the crosslinked network with $x_{\rm f} = 1.0$ exhibits the highest loading force and the high extent of bond breaking ($\mathcal{B}$). In contrast, the networks with $x_{\rm f} = 0.5$ and $x_{\rm f} = 0.2$ show significantly lower loading forces and corresponding bond breaking percentages ($\mathcal{B}$), as highlighted in Fig.~\ref{fig: Tetra_tri_loading_bond_breaking}(a–c) in red (right y-axis). As $x_{\rm f}$ decreases, the density of crosslinking bonds is reduced, resulting in less rigidity to resist deformation.

\subsection {Structural stiffness evaluation}
\label{sec: setwork stiffness}

The phenomenon of visco-elasticity manifests in materials such as polymers, hydro-gels, elastomers, and microgels, particularly under conditions of small deformation. It is a complex interplay between viscous and elastic properties, allowing these materials to exhibit time-dependent stress and strain behavior\cite{ain2024mechanical}. Even at a considerable amount of bond breaking happens in the crosslinked network while indentation, and thus is expected to follow elastic-plastic deformation. Previous studies have shown that the depth-sensing material can be investigated by its loading and unloading behaviour\cite{oliver2004measurement}. 

Analyzing the loading and unloading behavior of a crosslinked network while indenting, the properties like hardness, stiffness, and the effective elastic modulus can be calculated \cite {mauryaWCP, ain2024role}. The force displacement curve for the crosslinked network with different fractions of crosslinked monomers ($x_{\rm f}$), are shown in Figure~S1a, S1b, and S1c in the Supporting Information. From the force displacement data that we observed, the peak force during loading and slope of the unloading curve with different $x_{\rm f}$ are shown in Table \ref{tab: stiffness}. 

Oliver and Pharr have proposed a theory that relates the material stiffness to the mechanics observed during the unloading cycles. It is generally known as effective elastic modulus ($E_{\rm eff}$) as expressed in Eq.~\ref{eq: Effective elastic modulus}. 

\begin{equation}
    \label{eq: Effective elastic modulus}
    E_{\rm eff} = \frac{S}{2}\sqrt{\frac{\pi}{A_{\rm c}}}
\end{equation}

Here, $S$ is the slope of unloading curve and $A_{\rm c}$ is the area of contact for an indenter as a function of contact depth $d_{\rm c}$, as described in Eq.~\ref{eq: area of contact}, where contact depth can be calculated by using Eq.~\ref{eq: Contact depth}.

\begin{equation}
\label{eq: area of contact}
     A_{\rm c} = 2\pi r_{\rm ind}d_{\rm c} - \pi d_{\rm c}^{2}
\end{equation}

\begin{equation}
\label{eq: Contact depth}
     d_{\rm c} = d_{\rm max} - 0.75 F_{\rm max}/S
\end{equation}

This idea was originally created for the Berkovich contact, as explained in an article on measuring techniques\cite{oliver2004measurement}. This hypothesis is commonly used in nanoscale and microscale indentation investigations.
The network stiffness data is compiled in Table~\ref{tab: stiffness}. From the $E_{\rm eff}$ data in Table~\ref{tab: stiffness}, it is clearly shown that the effective elastic modulus ($E_{\rm eff}$) improved with increasing the $x_{\rm f}$ of the crosslinked network. Crosslinking restricts the movement of polymer chains, reducing their ability to deform under stress, making the material stiffer and more elastic due to the reduced freedom of the chain segments between crosslinks.\\
 
 $E_{\rm eff}$ is not valid only in the case of, when contact depth $d_{\rm c}$ will be more than the indenter diameter, i.e $ d_{\rm c} > 2r_{\rm ind}$.

\begin{table}[]
    \centering
    \begin{tabular}{|c|c|c|c|c|c|c|c|}
    \hline
    
       $r_{\rm ind}$ & $x_{\rm f}$  & $d_{\rm max}$ & $F_{\rm max}$ & $S $ & $ d_{\rm c}$ & $ A_{\rm c} $ & $E_{\rm eff} $ \\
       \hline
      $[\sigma]$ &   & $[\sigma]$ & $[k_{\rm B}T/\sigma]$ & $[~k_{\rm B}T/\sigma^{2}]$ & $ [\sigma]$ & $ [\sigma^{2}] $ & $[~k_{\rm B}T/\sigma^{3}]$ \\
       \hline

          $8.0$& $1.0$ & $13.50$ & $14342.0$ & $2464.7$ & $9.13$ & $197.08$ & $155.61$ \\
         
          &$0.5$ & $13.50$ & $2202.7$ & $354.59$ & $8.84$ & $198.92$ & $22.28$ \\

           &$0.2$ & $14.50$ & $517.61$ & $69.56$ & $8.91$ & $198.48$ & $4.34$ \\
        
         \hline
         
    \end{tabular}
    \caption{This table presents the effective elastic modulus ($E_{\rm eff}$), peak loading force ($F_{\rm max}$), and unloading slope ($S$) as functions of $x_{\rm f}$ for crosslinked polymer networks.}
    \label{tab: stiffness}
\end{table}

\subsection{Adhesion analysis}
\label{Analysis of force of adhesion}

\begin{figure*}[t]
\centering
\hfill
\includegraphics[width=1.0\textwidth,angle=0]{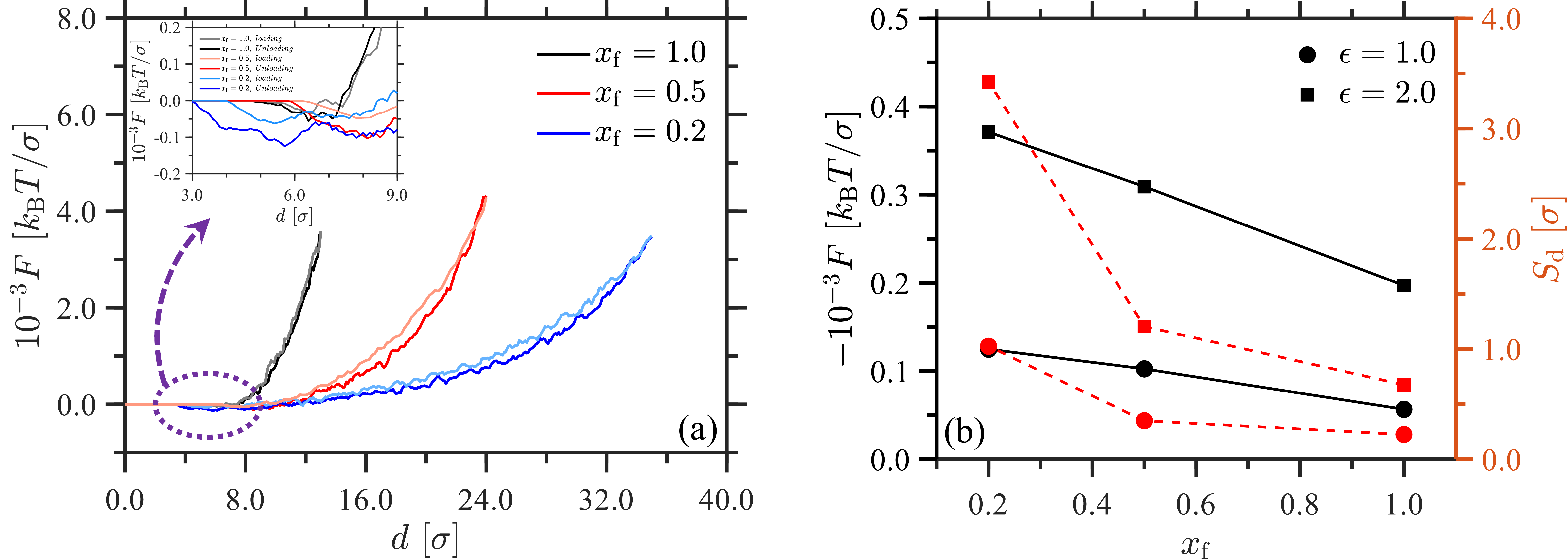}
\caption{(a) Loading and unloading force versus indentation depth. The magnified inset in Figure (a) highlights the negative pull-off force observed during indenter unloading for $\epsilon = 1.0$ with cut-off distances $r_{\rm c} = 2.5\sigma$. Figure (b) shows the pull-off force (left y-axis) and the corresponding separation distance $S_{\rm d}$ (right y-axis) during the unloading of the indenter. Data are shown for tetrafunctional networks with different fractions of crosslinked monomers. The indenter approach velocity was $v = 0.005\sigma/\tau$, and the indenter radius was $8.0\sigma$\\
Note: Separation distances ($S_{\rm d}$) were calculated as the difference in indentation depth between the point where the loading force begins and the point where the unloading force reaches zero.}
\label{fig: Pull off force}
\end{figure*}

For adhesion analysis, the indentation test was conducted on crosslinked networks with different fractions of the crosslinked monomers ($x_{\rm f}$) As illustrated in Fig.~\ref{fig: Pull off force}(a), for all the simulations, the indentation was performed up to an indentation depth where no bond breaking occurred, purely elastic deformation. Beyond this depth, bond breaking initiates. Therefore, within the elastic regime, the loading and unloading curves overlap. From the zoomed-part inset in Fig.~\ref{fig: Pull off force}(a), it is clearly shown that the adhesion force (or pull-off force) during unloading increases as $x_{\rm f}$ of the crosslinked network decreases. This observation was made while keeping the interaction strength value ($\epsilon$) fixed at 1.0 in the attractive regime ($r_{\rm c} = 2.5\sigma$). Also, we calculate the separation distance $S_{\rm d}$~(Separation distances were calculated as the difference in indentation depth between the point where the loading force begins and the point where the unloading force reaches zero) while unloading increases with reducing $x_{\rm f}$, which primarily resulting increase in network flexibility and chain mobility. When the polymer network becomes less densely crosslinked, it behaves more like a soft material, that can deform more readily under mechanical stress. This increase in the network deformation allows for more real contact area and interaction of monomers with the counter surfaces, leading to stronger van der Waals or adhesive interactions in the attractive regime, thereby increasing the measured separation distance ($S_{\rm d}$) and pull off force. This behavior reflects the enhanced ability of a softness.

To further investigate the influence of interaction strength on both the adhesion force and the separation distance ($S_{\rm d}$), we systematically increased the epsilon ($\varepsilon$) value from 1.0 to 2.0 (as illustrated in Figure~S2 in the Supporting Information). The results demonstrate a noticeable increase in both the adhesion force and the separation distance within the attractive regime, as depicted in Fig.~\ref{fig: Pull off force}(b). This behavior can be attributed to the presence of stronger intermolecular forces between the contacting surfaces, which require more amount of energy and mechanical displacement to overcome during the indenter unloading process.


\begin{figure*}[t]
\hfill
\includegraphics[width=1.0\textwidth,angle=0]{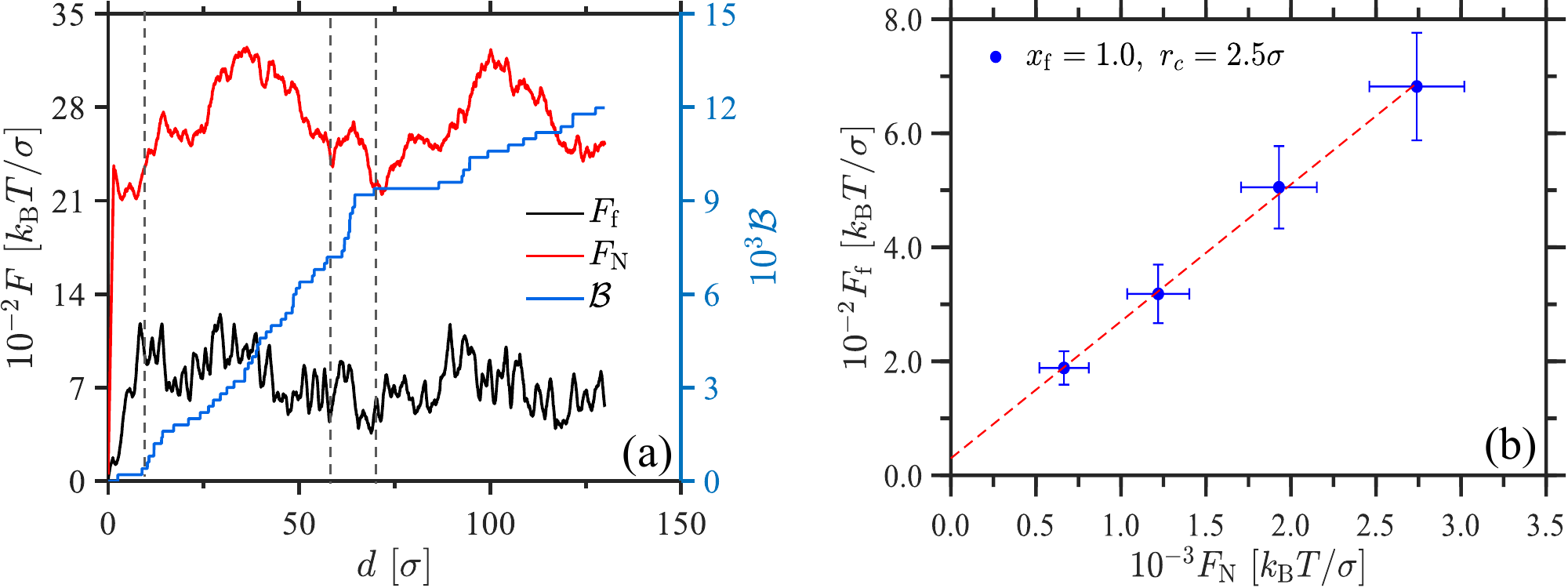}
\caption{Part (a) presents the normal and friction forces (left y-axis) alongside the corresponding bond-breaking percentage ($\mathrm{B}$) (right y-axis, sky-blue color) plotted against sliding distance at an indentation depth of $4\sigma$. The dashed line marks the point of maximum bond breaking. Part (b) depicts the average normal force versus friction force data at various indentation depths during friction, dashed red line representing the linear regression of the corresponding data sets. These data correspond to a tetrafunctional crosslinked network with $x_{\rm f} = 1.0$, an indenter radius was $r_{\rm ind} = 8.0\sigma$, and a sliding velocity was  $0.005\sigma/\tau$.}
\label{fig: T4E1rc2.5_FNB@d4}
\end{figure*}

\subsection{Determination of Coefficient of Friction}
\label{subsection: CoF calculation}

The friction force differs from a conventional applied force in the Newtonian sense: while an applied force acts externally to cause acceleration, friction is a contact force that arises at the interface between surfaces and always acts to oppose relative motion between them. Therefore, friction can be considered as a reaction force rather than an action force\cite{amontons1999resistance,blau2001significance}. The coefficient of friction (CoF) between two rubbing surfaces is defined by Eq.~\ref {eq: CoF equation}.

\begin{equation}
\label{eq: CoF equation}
    F_{\rm f} = \mu  F_{\rm N} + F_{\rm adh}
\end{equation}

Note: Here, $F_{\rm f}$ is the friction force, $\mu$ is the coefficient of friction (CoF), $F_{\rm N}$ is the normal force and $F_{\rm adh}$ is the adhesion force.

For determining the friction force, we removed the repulsive walls from the system and applied periodic boundary conditions in $x$, $y$, and $z$ directions and ran the simulations using the micro-canonical ensemble with the system temperature of $T = 1\epsilon/~k_{\rm{B}}$. The data was recorded by the system with indenter sliding velocity $0.005\sigma/\tau$ and for sliding distance ($S$) $135\sigma$, for different fractions of crosslinked monomers ($x_{\rm f}$). The friction force ($F_{\rm f}$) and normal force ($F_{\rm N}$), along with the corresponding bond-breaking percentage ($\mathcal{B}$) with $x_{\rm f} = 1.0$ at $4\sigma$ depth of indentations are shown in Fig.~\ref{fig: T4E1rc2.5_FNB@d4}(a), the dashed line indicates the location of sliding distance at which maximum bond breaking occurs, leading to a sharp drop in both the normal force and friction force. Force response depends upon the crosslinking bond density of the network ($x_{\rm f}$). If indentation depth increases while sliding, the friction force and normal force increase.

Furthermore, to understand the sudden increase and decrease in the frictional and normal forces ($\delta F$) during sliding, we quantified the percentage of bond breaking, denoted as $\mathcal{B}$. As clearly shown in Fig.~\ref{fig: T4E1rc2.5_FNB@d4}(a) (sky-blue line, right y-axis), the bond-breaking percentage is significantly higher during the first cycle of friction (from 0 to 67.5$\sigma$) compared to the second cycle (from 67.5 to 135.0$\sigma$). This behavior can be attributed to the structural rearrangement and relaxation of the monomers during the first sliding cycle. In the first cycle, during sliding of the indenter, bonds are broken due to interfacial shear. As a result, the system undergoes relaxation, leading to fewer bond breakages in the second sliding cycle. 

We performed four simulations under controlled conditions, each with a different indentation depth of the indenter into the substrate: $1.0\sigma$, $2.0\sigma$, $3.0\sigma$, and $4.0\sigma$. The simulation results show that as the indentation depth increases, the friction force and corresponding normal force also increase. The bond-breaking percentage increases with indentation depth during friction with $x_{\rm f} = 1.0$ only. Additionally, no bond breaking was observed during sliding in the crosslinked networks with $x_{\rm f}$ 0.5 and 0.2.

To analyze this, we considered the friction and normal force data over the same sliding distance, excluding the static region before the first dashed line as shown in Fig.~\ref{fig: T4E1rc2.5_FNB@d4}(a). Then we performed a linear regression on this friction and normal force data, and the slope and intercept of the resulting regression line (indicated by the red dashed line in Fig.~\ref{fig: T4E1rc2.5_FNB@d4}(b)) represent the average coefficient of friction (CoF) and adhesion force ($F_{\rm adh}$), respectively. In this case, the CoF was found to be 0.2401, and the adhesion force was 29.98 $k_{\rm B}T/\sigma$.

\begin{figure}[h!]
\centering
\includegraphics[width=0.5\textwidth,angle=0]{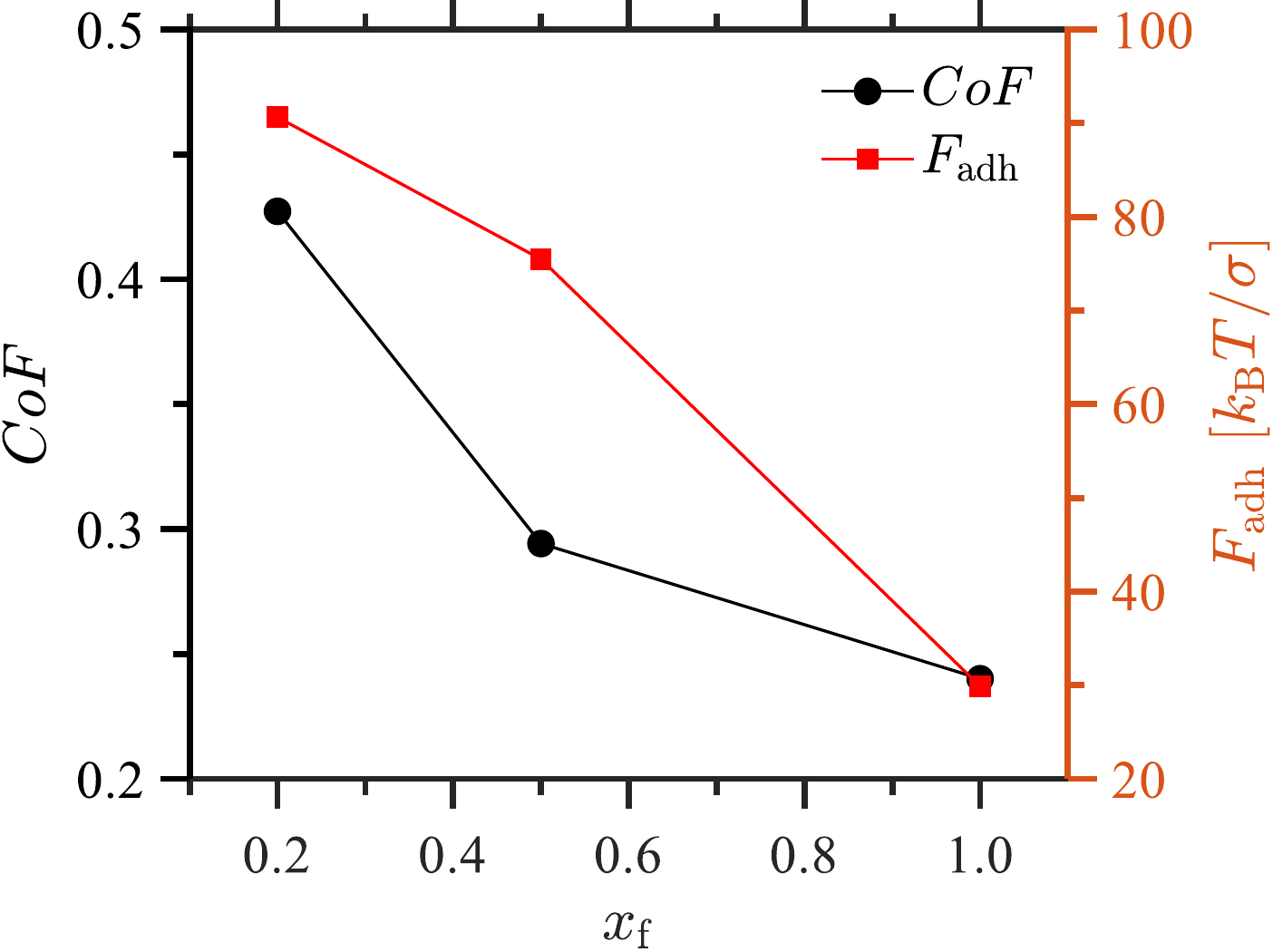}
\caption{Coefficient of friction (CoF) (left y-axis, black color) and average adhesion force $F_{\rm adh}$ during friction (right y-axis, red color) as a function of $x_{\rm f}$.} 
\label{fig: CoF_vs_monomer fractions}
\end{figure}

In order to investigate the influence of the fraction of crosslinked monomers ($x_{\rm f}$) on the CoF and adhesion force ($F_{\rm adh}$), we conducted simulations for different $x_{\rm f}$, shear velocity was kept constant at $0.005\sigma/\tau$ for all simulations. as shown in Fig.~\ref{fig: CoF_vs_monomer fractions} and Figure~S3 in the Supporting Information. It is clearly shown that the slope of the regression line (CoF) and the intercept on the y-axis (adhesion force) decrease with increasing $x_{\rm f}$. The conformational ranges of the crosslinked polymer network are determined by the molecular mobility of the chains within the network and connecting the crosslinking junctions\cite{bhuyan2007effect}. As $x_{\rm f}$ increases, a greater number of monomers are bonded with the neighboring monomers, which reduces the chain mobility within the network. This leads to a stiffer network. A reduction in chain mobility leads to lower adhesive interactions at the sliding interface\cite{alazemi2016experimental,mok2018characterisation}. Furthermore, stiffer materials deform less under load, resulting in reduced real contact area with the indenter, which contributes to a decrease in CoF.

\section{Conclusions}
\label{sec: Conclusion}

To know the mechanical and tribological behavior of the crosslinked network, we used large-scale molecular dynamics simulation approach based on a generic model (chemically independent). This approach has various advantages over traditional tribological and tensile indentation methods, particularly used for knowing the effect of local mechanical response. However, because the bond breaking and network configuration are easily recognized, deformations at the monomer level can be discovered and analyzed. The results showed an interesting mechanical response in the form of a force-drop, which resembles the avalanches observed in glassy types of materials. The force drops in cross-linked polymer networks are associated with bond breaking. In the current study, quartic potential parameters were utilized to create an environment that is very reminiscent of amine-based epoxies. At the same time, FENE bonds are used to create an environment that is highly reminiscent of the softer bond created by alkane-based linkers. we find a more in-depth analysis of the mapping techniques from atomistic to generic. It is also important to highlight how mechanical indentation and friction are performed on cross-linked systems. The fact that these systems produce a force response that varies with depth is one of the intriguing things about them, and the network stiffness increases with increase in the fraction of crosslinked monomers. Additionally, it is observed that the force of adhesion and coefficient of friction (CoF) increase as the fraction of crosslinked monomers decreases.

\section*{Supporting Information}
This file contains additional data on structural stiffness evaluation and adhesion analysis.

\section*{Conflicts of interest}

The authors declare no competing interests.

\section*{Data Availability}

The scripts and the data associated
with this research are available upon reasonable request
from the corresponding author(s)

\section*{Acknowledgements}
A.K. and M.K.M. thank the National Supercomputing Mission (NSM) facility PARAM
Sanganak at IIT Kanpur, where all the simulations were
performed. M.K.S. thanks the funding support provided by IIT Kanpur under the initiation grant scheme.


%

\end{document}